\documentclass[twocolumn]{webofc}
\usepackage[varg]{txfonts}   
\usepackage{booktabs}
\woctitle{FUSION23}
\makeatletter
\def\NAT@def@citea{\def\@citea{\NAT@separator}}
\makeatother 

\begin{document}
\title{Pauli energy contribution to nucleus-nucleus interaction}
%
%

\author{\firstname{A.S.} \lastname{Umar}\inst{1}\fnsep\thanks{\email{umar@compsci.cas.vanderbilt.edu}} \and
        \firstname{K.} \lastname{Godbey}\inst{2}\fnsep\thanks{\email{kylegodbey@gmail.com}} \and
        \firstname{C.} \lastname{Simenel}\inst{3}\fnsep\thanks{\email{cedric.simenel@anu.edu.au}}
}

\institute{Department of Physics and Astronomy, Vanderbilt University, Nashville, TN 37235, USA
\and
           Facility for Rare Isotope Beams, Michigan State University, East Lansing, Michigan 48824, USA
\and
           Department of Fundamental and Theoretical Physics and Department of Nuclear Physics and Accelerator Applications, Research School of Physics, The Australian National University, Canberra ACT  2601, Australia
          }

\abstract{
The investigation delves into understanding how the Pauli exclusion principle
influences the bare potential between atomic nuclei through the application of
advanced theoretical methodologies. Specifically, the application of the novel
Frozen-Hartree-Fock (DCFHF) technique is employed. The resulting potentials
demonstrate a noticeable repulsion at short distances, attributed to the effects
of the Pauli exclusion principle. To account for dynamic phenomena, such as
nucleon transfer processes, the density-constrained time-dependent Hartree-Fock
(DC-TDHF) method is utilized. This approach integrates isovector contributions
into the potential, shedding light on their influence on fusion reactions.
Notably, the inclusion of isovector effects leads to a reduction or enhancement
in the inner part of the potential, suggesting a nuanced role of transfer in the fusion process.
}
\maketitle
\section{Introduction} 
\label{intro}
The Pauli exclusion principle plays a crucial role as a building block of many-body quantal
systems comprised of fermions. It also induces a "Pauli repulsion" in the interaction between di-nuclear systems.
It has been shown that~\cite{simenel2017} the Pauli repulsion widens the nucleus-nucleus
potential barrier, thus hindering sub-barrier fusion.
We investigate the proton and neutron contributions to the Pauli repulsion, both in the bare potential
neglecting shape polarization and transfer between the reactants, as well as in the dynamical potential obtained
by accounting for such dynamical rearrangements.
As the basis of our study we utilize the Pauli kinetic energy (PKE) obtained by studying the nuclear
localization function (NLF)~\cite{reinhard2011}. Recently this approach has been generalized to incorporate all of the dynamical
and time-odd terms present in the nuclear energy density functional~\cite{li2020}. This approach is employed in the density
constrained frozen Hartree-Fock (DCFHF) and in the density constrained time-dependent Hartree-Fock (DC-
TDHF) microscopic methods.
The PKE spatial distribution shows that a repulsion occurs in the neck between the nuclei when they
first touch. Inside the barrier, neutrons can contribute significantly more to the Pauli repulsion in neutron-rich
systems. Dynamical effects tend to lower the Pauli repulsion near the barrier. Proton and neutron dynamical
contributions to the PKE significantly differ inside the barrier for asymmetric collisions, which is interpreted as
an effect of multinucleon transfer.
The PKE is shown to make a significant contribution to nuclear interaction potentials. Protons and
neutrons can play very different roles in both the bare potential and in the dynamical rearrangement. Further
microscopic studies are required to better understand the role of transfer and to investigate the effect of pairing
and deformation~\cite{umar2021}.

\section{Formalism}
\label{sec-1}
\subsection{Microscopic Methods}
To explore the influence of the Pauli energy in heavy-ion fusion reactions, our
methodology incorporates microscopic techniques to calculate the interactions
between nuclei. We base our approach on the energy density functional (EDF) to compute
nucleus-nucleus potentials.
The computation of the bare potential begins with the assumption of frozen
ground-state densities for each nucleus, derived from the Hartree-Fock (HF)
mean-field approximation. This method leverages the Skyrme EDF for both the HF
calculations and the potential computation, ensuring consistency without
introducing additional parameters.
This potential is derived from the spatial integral of the energy density as a
function of the nuclear densities
\begin{equation}
V_\mathrm{FHF}(\mathbf{R})=\int d\mathbf{r}\;\mathcal{H}\left[\rho_1(\mathbf{r})+\rho_2(\mathbf{r}-\mathbf{R})\right]- E[\rho_1] -E[\rho_2]\,.
\label{eq:frozen}
\end{equation}
Here, Pauli exclusion principle's effects are initially set
aside, except for those emanating from the exchange terms in the effective
interaction, leading to the formulation of the conventional 
FHF potential~\cite{denisov2002,skalski2007,washiyama2008,simenel2008,guo2012,vophuoc2016}.

To incorporate the Pauli repulsion into the bare potential, we adopt
the DCFHF method~\cite{simenel2017}. This approach includes the Pauli exclusion principle
explicitly by allowing the reorganization of single-particle states within the
combined nuclear density to achieve a minimum energy configuration. This
reorganization results in a unique Slater determinant, with the HF minimization
performed under constraints that maintain the local densities of protons and
neutrons unchanged
\begin{equation}
\delta \left\langle \ H - \sum_{q=p,n}\int\, d\mathbf{r} \ \lambda_q(\mathbf{r})  \left[\rho_{1_q}(\mathbf{r})+\rho_{2_q}(\mathbf{r}-\mathbf{R})\right] \ 
\right\rangle = 0\,.
\label{eq:var_dens}
\end{equation}
The DCFHF method yields bare potentials that acknowledge the Pauli repulsion,
effectively widening the fusion barrier and generating a potential pocket at
closer distances not observed in FHF potentials
\begin{equation}
V_{\mathrm{DCFHF}}(R)=\langle\Phi(\mathbf{R}) | H | \Phi(\mathbf{R}) \rangle-E[\rho_1]-E[\rho_2]\,.
\label{eq:vr}
\end{equation}
This difference enables a
comparative study of the Pauli principle's impact on frozen nuclear densities.

For dynamic nuclear interactions, we turn to the time-dependent Hartree-Fock
(TDHF) calculations. These calculations consider the rearrangement of densities
at the mean-field level, affected by couplings to vibrational and rotational
modes, as well as nucleon transfer mechanisms. The potentials derived from TDHF
calculations, therefore, reflect both dynamical effects and the Pauli exclusion
principle. To delve deeper into these dynamics, we employ the
Density-Constrained TDHF (DC-TDHF) method, directly using the densities from
TDHF system evolution while applying the same constraint procedure as in DCFHF
\begin{equation}
V_{\mathrm{DC-TDHF}}(R)=\langle\Phi(\mathbf{R(t)}) | H | \Phi(\mathbf{R(t)}) \rangle-E[\rho_1]-E[\rho_2]\,.
\label{eq:vrdctdhf}
\end{equation}
The DC-TDHF approach~\cite{umar2006b,washiyama2008,guo2012,washiyama2020,scamps2019b} allows for a nuanced understanding of microscopic phenomena
related to the Pauli principle, including orbital splitting and its attractive
or repulsive contributions to the potential. However, it's crucial to note the
inherent limitations of TDHF-based methods, such as their inability to account
for many-body tunneling effects, which leaves the energy dependency of the
potential at sub-barrier energies as an open question. Despite these
limitations, the application of DC-TDHF potentials in calculating near- and
sub-barrier fusion cross-sections has demonstrated considerable success in
aligning with experimental outcomes, showcasing the method's efficacy in
capturing the complex interplay of nuclear forces at play in heavy-ion fusion
reactions.

\subsection{Localization function and PKE}
\begin{figure}[!htb]
\centering
  \includegraphics*[width=4cm]{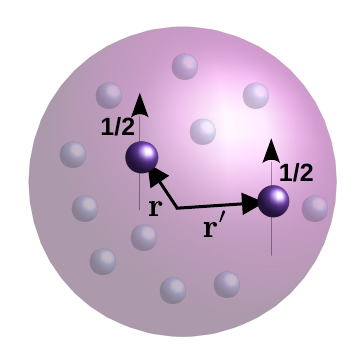}
  \caption{\protect Schematic depiction of two nucleons at $\mathbf{r}$ and $\mathbf{r'}$ with
  spin up along the $z$-axis entering Eq.~(\ref{conditionalProb}).}
  \label{fig1}
\end{figure}

This section delves into the concept of the localization function
for nuclear systems~\cite{reinhard2011,li2020} and
its connection to kinetic energy and the Pauli exclusion principle. 
The conditional probability for finding a nucleon at
$\mathbf{r'}$, shown in Fig.~\ref{fig1},  when we know with certainty that another nucleon with the same
spin and isospin is at $\mathbf{r}$ is proportional to
\begin{equation}
	\label{conditionalProb}
	R_{qs}(\mathbf{r},\mathbf{r}')=
	\frac{\rho_q(\mathbf{r}s,\mathbf{r}s)\rho_q(\mathbf{r}'s,\mathbf{r}'s)-|\rho_q(\mathbf{r}s,\mathbf{r}'s)|^2}{\rho_q(\mathbf{r}s,\mathbf{r}s)}\,,
\end{equation}
where $\rho_q$ is the component of the one-body density matrix with isospin $q$. 

The short-range behavior of $R_{qs}$ can be 
obtained using techniques similar to the local density approximation~\cite{reinhard2011,li2020}. The
leading term in the expansion yields the localization measure
\begin{equation}\label{eq:prob_D}
    D_{qs_{\mu}} = \tau_{qs_{\mu}}-\frac{1}{4}\frac{\left|\boldsymbol{\nabla}\rho_{qs_{\mu}}\right|^2}{\rho_{qs_{\mu}}}
    -\frac{\left|\mathbf{j}_{qs_{\mu}}\right|^2}{\rho_{qs_{\mu}}}\,,
\end{equation}
where the densities and
currents are given in their most unrestricted form \cite{li2020} for
$\mu$ axis denoting the spin-quantization axis.
This measure is the most general form that is appropriate for deformed nuclei and without assuming
time-reversal invariance, thus also including the time-odd terms important in applications such as
cranking or TDHF. 
We can visualize the NLF defined from the localization measure in Eq.~(\ref{eq:prob_D}).
It is advantageous to normalize the localization measure to the interval $[0,1]$ using~\cite{liu2019b,li2020}
\begin{equation}
  \mathcal{D}_{qs_\mu}(\mathbf{r})=\frac{D_{qs_\mu}(\mathbf{r})}{\tau_{qs_\mu}^{\mathrm{TF}}(\mathbf{r})}\,, 
\end{equation}
where the normalization 
$\tau_{qs_\mu}^{\mathrm{TF}}(\mathbf{r})=\frac{3}{5}\left(6\pi^2\right)^{2/3}\rho_{qs_\mu}^{5/3}(\mathbf{r})$ is the Thomas-Fermi kinetic density.
The NLF can then be represented by
\begin{equation}
  {C}_{qs_\mu}(\mathbf{r})=\left[1+\mathcal{D}_{qs_\mu}^{2}\right]^{-1}\,.
  \label{eq:NLF}
\end{equation}

To calculate the Pauli kinetic energy the expression in Eq.~(\ref{eq:prob_D}) can be
dissected into two components. The last two terms are the kinetic density for
a complex valued single particle state of a given spin $s$ and
isospin $q$. The first term represents the von
Weizsacker kinetic-energy density, and together, these provide the kinetic density~~\cite{li2020}
\begin{equation}
	\tau_{qs}^{\rm s.p.}= \frac{1}{4}\frac{\left|\boldsymbol{\nabla} \rho_{qs}\right|^2}{\rho_{qs}} + \frac{\left|\mathbf{j}_{qs}\right|^2}{\rho_{qs}}\,.
    \label{eq:tsp}
\end{equation}
Thus, one can write
\begin{equation}
	D_{qs}=\tau_{qs}-\tau_{qs}^{\rm s.p.}\,.
\end{equation}
This equation establishes $D_{qs}$ as the difference between the total kinetic
energy density and the kinetic energy density for a single-particle state, being
zero for a single nucleon system. $D_{qs}$ directly measures the additional kinetic
density brought about by the Pauli exclusion principle, which prevents two
fermions (such as protons or neutrons in a nucleus) from occupying the same
quantum state.
The total PKE for a nuclear system can be obtained by integrating
\begin{equation}
	E_{qs}^\mathrm{P}=\frac{\hbar^2}{2m}\int d^3r\; D_{qs}(\mathbf{r})\,.\label{eq:EPauli}
\end{equation}
Since the PKE for a single nucleus does not contribute to the Pauli repulsion for the
nucleus-nucleus potential, we can define the difference
\begin{equation}\label{pke-1}
	\Delta E_{q\mu}^{\mathrm{P(F)}}(R)=\frac{\hbar^2}{2m}\sum_{s_\mu}\int d^3r\; \left[ D_{qs_{\mu}}^\mathrm{DCFHF}(\mathbf{r},R)- D_{qs_{\mu}}^\mathrm{FHF}(\mathbf{r},R)\right]\,,
\end{equation}
where we have subtracted the contribution of the PKE from the FHF approach
and summed
over the spin-up and spin-down components for a given spin
projection axis $\mu$.
Indeed, the latter uses the same frozen density as DCFHF, but it neglects the Pauli exclusion principle between nucleons of different nuclei. 
The notation $\mathrm{P(F)}$ stands for ``Pauli in the Frozen approximation''.
A similar expression can be constructed by subtracting DC-TDHF and DCFHF contributions to identify
the dynamical contribution to PKE.

\section{Results}
\begin{figure}[!htb]
\centering
  \includegraphics[width=0.5\columnwidth]{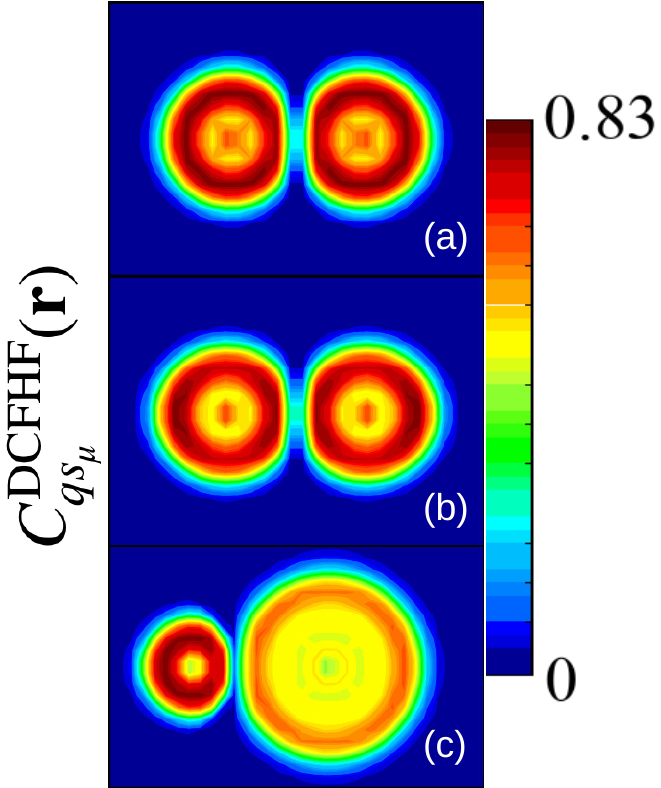}
  \caption{\protect Plotted are the NLF's, obtained with the DCFHF method
for spin up along the $z$-axis, corresponding to (a) $^{40}$Ca+$^{40}$Ca, 
(b) $^{48}$Ca+$^{48}$Ca, and (c) $^{16}$O+$^{208}$Pb systems at $R\approx$ 11~fm.}
  \label{fig2}
\end{figure}

One of the interesting results is the observation of Pauli repulsion directly through NLF. In
Fig.~\ref{fig2} we show the stacked plot of NLF's for neutron distributions, obtained with the DCFHF method
for spin up along the $z$-axis, corresponding to (a) $^{40}$Ca+$^{40}$Ca, 
(b) $^{48}$Ca+$^{48}$Ca, and (c) $^{16}$O+$^{208}$Pb. 
The NLF for protons is very similar to that of the neutrons.
The approximate distance between the nuclei is about 11~fm. 
We note the effect of the Pauli repulsion on the single-particle states resulting from the
DCFHF calculation, concentrating in the region of touching surfaces.
We should remember that
since the densities are frozen the corresponding density plots will not show any Pauli
effects.

\begin{figure}[!htb]
\centering
  \includegraphics[width=0.95\columnwidth]{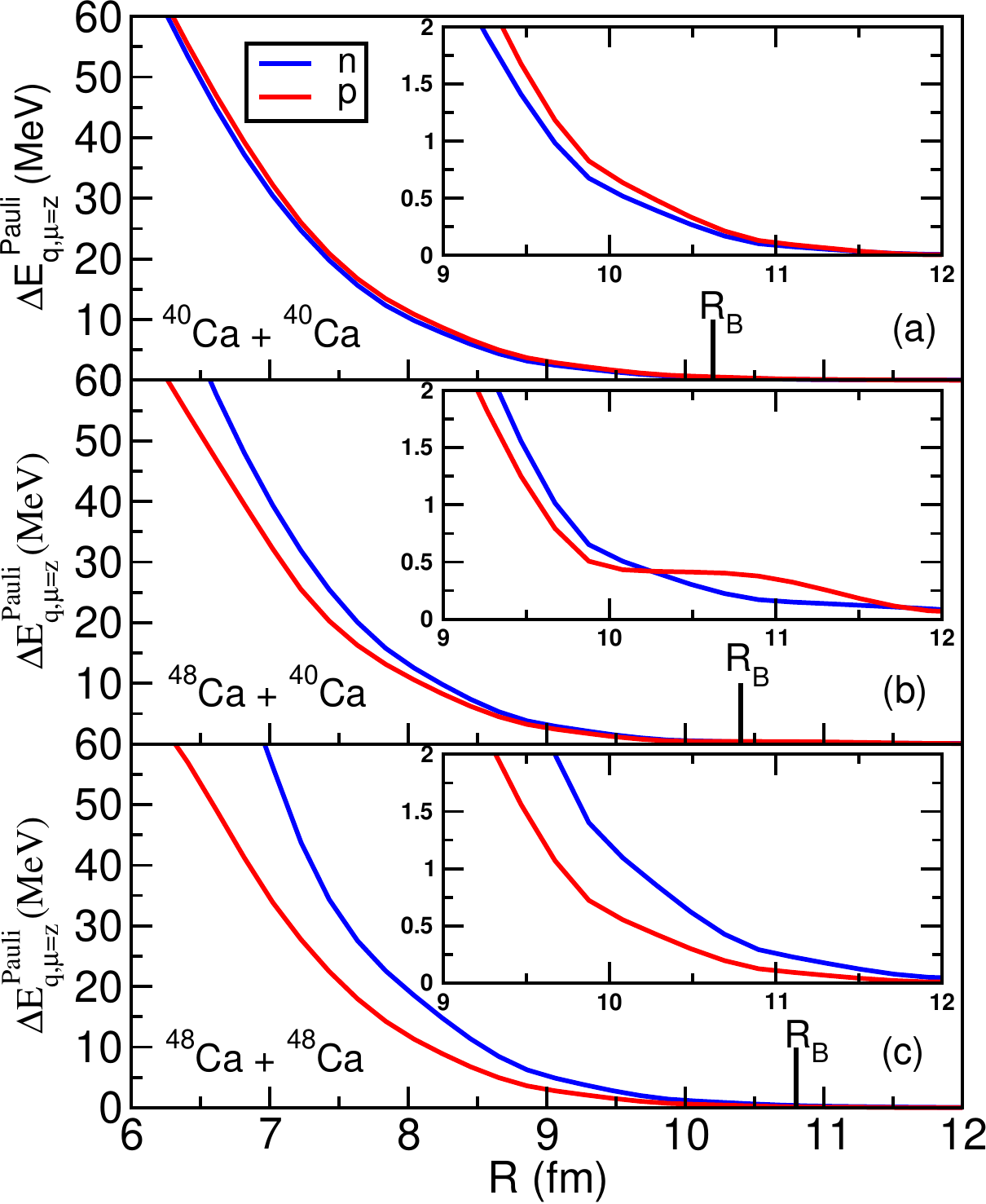}
  \caption{\protect Neutron and proton contributions to the  Pauli repulsion
           from Eq.~(\ref{pke-1}) in the frozen approximation
	   for the $^{40,48}$Ca$+^{40,48}$Ca systems. The insets focus on the
	   barrier top region.}
  \label{fig3}
\end{figure}

The contributions of protons and neutrons to Pauli repulsion, calculated
using Eq.~(\ref{pke-1}) under the assumption of frozen nuclear densities, are
depicted in Fig.~\ref{fig3} for the $^{40,48}$Ca$+^{40,48}$Ca systems. 
A pronounced increase in Pauli repulsion is
observed inside the barrier for all examined nuclear systems.
The kinetic energy due to the Pauli principle becomes significantly large
at shorter distances between the ions. This phenomenon underlies the formation of a
potential well at close separations, suggesting that the increase in PKE
counterbalances the rapid decrease in the interaction potentials derived from
static (frozen) nuclear densities.

In the system of $^{40}$Ca$+^{40}$Ca, where the number of protons equals the number of 
neutrons ($N = Z$), the contributions from both protons and neutrons to the Pauli
energy are almost identical. 
For the systems that do not have extended neutron skins the protons interact earlier
due to the Coulomb interaction, which leads to an early increase for proton PKE.
While the proton contribution remains
relatively constant across the $^{40,48}$Ca$+^{40,48}$Ca systems, the Pauli contribution
noticeably strengthens with an increase in the neutron number within the
barrier. For example, at a separation of $R\approx 9$~fm in the $^{48}$Ca$+^{48}$Ca system,
corresponding to the inner turning point of a quantum tunneling path at roughly
0.9 times the barrier potential ($V_B$), the contribution is predominantly
exerted by neutrons. Consequently, the neutron contribution to the Pauli energy
significantly surpasses the proton contribution.

\begin{figure}[!htb]
\centering
  \includegraphics[width=0.95\columnwidth]{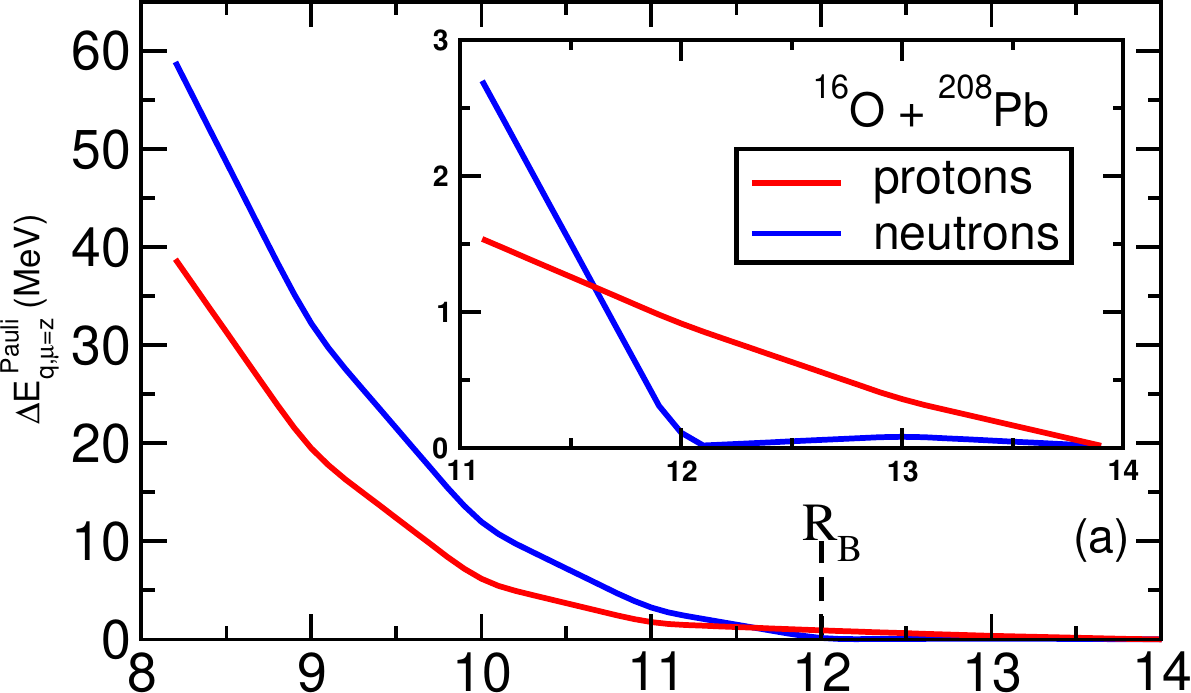}
  \caption{\protect Neutron and proton contributions to the  Pauli repulsion
           from Eq.~(\ref{pke-1}) in the frozen approximation
	   for the $^{16}$O+$^{208}$Pb system. The inset focuses on the
	   barrier top region.}
  \label{fig3b}
\end{figure}

A similar trend is observed in the $^{16}$O+$^{208}$Pb system shown in Fig.~\ref{fig3b}, where the neutron contribution
is roughly double that of the proton force contribution inside the barrier. This pattern may
be attributed to the formation of neutron skins in neutron-rich nuclei, which
results in neutrons interacting first near the barrier's edge. Therefore, the
Pauli exclusion principle predominantly affects neutrons in such configurations.
Consequently, one may expect a fusion hindrance due to the
Pauli exclusion principle in neutron-rich systems, in particular
at sub-barrier energies.

In the realm of frozen density frameworks, the DCFHF approach incorporates
antisymmetrization principles to yield a refined prediction of the Pauli Kinetic
Energy (PKE) at a static, mean-field level. Nonetheless, a dynamic examination
of nuclear reactions reveals that the densities of interacting nuclei are
dynamic, engaging with each other prior to surpassing the barrier peak. This
interaction leads to shape polarization and potential nucleon exchange between
the fragments, while adhering to the Pauli exclusion principle. Within density
functional theory (DFT), such dynamics are modeled through the TDHF (or TDDFT)
evolution. The DC-TDHF method calculates the nucleus-nucleus potential based on
these time-evolving densities, allowing for a nuanced comparison of Pauli
repulsion effects under static and dynamic conditions, thereby enhancing our
understanding of Pauli kinetic energy's dynamic development. To this end we
can define a similar measure
\begin{equation}\label{pke-D}
	\Delta E_{q\mu}^{\mathrm{P(D)}}(R)=\frac{\hbar^2}{2m}\sum_{s_\mu}\int d^3r\; \left[ D_{qs_{\mu}}^\mathrm{DC-TDHF}(\mathbf{r},R)- D_{qs_{\mu}}^\mathrm{DCFHF}(\mathbf{r},R)\right]\,.
\end{equation}

\begin{figure}[!htb]
\centering
  \includegraphics[width=0.95\columnwidth]{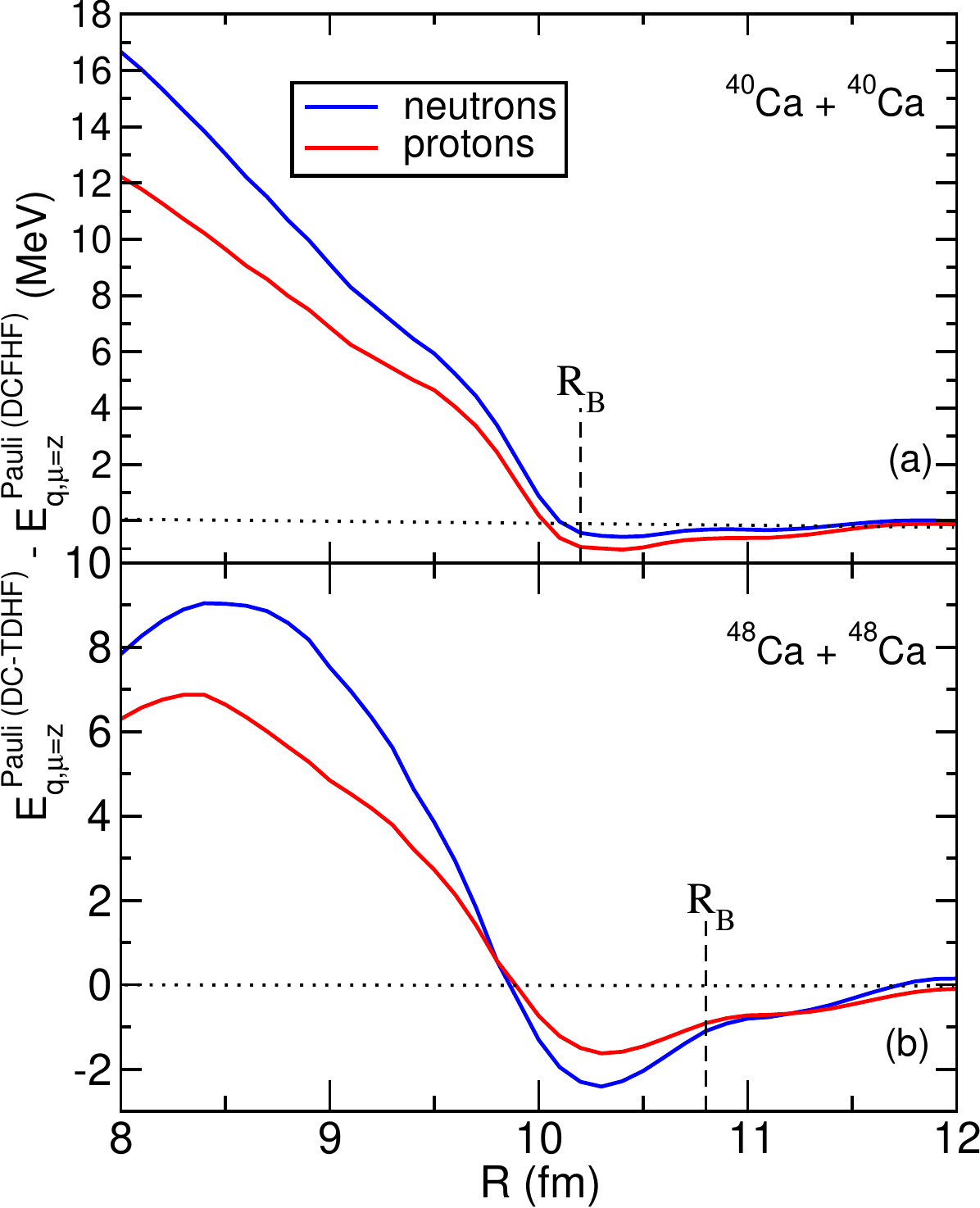}
  \caption{\protect Dynamical contributions to the Pauli repulsion computed from Eq.~(\ref{pke-D}) for the $^{40}$Ca$+^{40}$Ca and $^{48}$Ca$+^{48}$Ca) systems.}
  \label{fig4}
\end{figure}

\begin{figure}[!htb]
\centering
  \includegraphics[width=0.95\columnwidth]{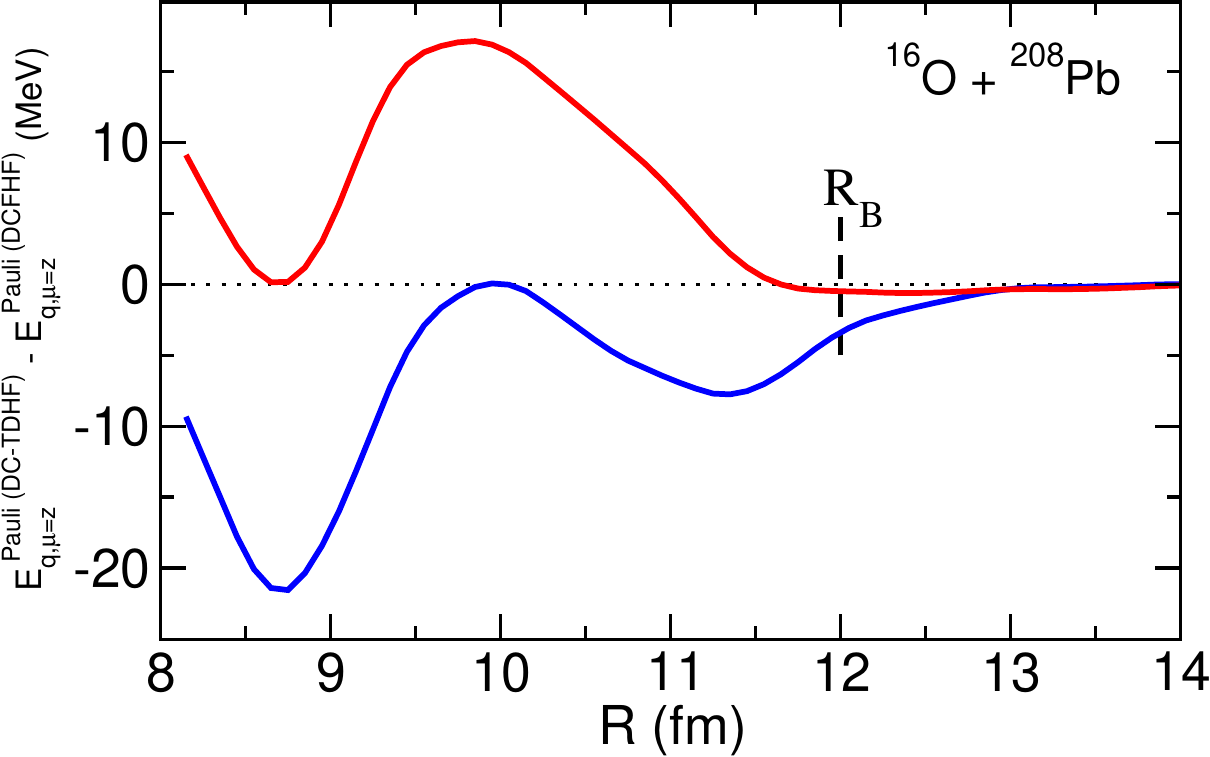}
  \caption{\protect Dynamical contributions to the Pauli repulsion computed from Eq.~(\ref{pke-D}) for the $^{16}$O+$^{208}$Pb system.}
  \label{fig5}
\end{figure}

Figures~\ref{fig4} and \ref{fig5} display the variance in Pauli energies as derived from
DC-TDHF and DCFHF methodologies relative to the internuclear distance, $R$. In
every instance, incorporating dynamics lowers the overall Pauli
repulsion across proton and neutron contributions near the barrier radius,
$R_B$. Given that TDHF formulations stem from a principle of stationary action,
which typically minimizes action (and since a rise in PKE would increase the
action), it's logical for the system to seek configurations that diminish PKE.
Closer internuclear distances, however, witness a surge in total PKE. It's
important to recognize that density configurations at smaller $R$ markedly differ
from static densities, making the PKE disparity at $R \leq R_B$ sometimes less
significant. Nonetheless, the dynamic-induced increase in Pauli repulsion at
close ranges might reflect a physical phenomenon, such as nucleon dynamical
collectivization, which elevates the PKE as collision partners coalesce.

Focusing on the differential roles of protons and neutrons in the dynamic
rearrangement of net PKE, depicted in Figs.~\ref{fig4} and \ref{fig5}, we find uniform
behavior in symmetric reactions ($^{40}$Ca$+^{40}$Ca and $^{48}$Ca$+^{48}$Ca), with neutrons
presenting a marginally stronger repulsion at shorter distances. Conversely, in
asymmetric collisions ($^{40}$Ca$+^{48}$Ca and $^{16}$O+$^{208}$Pb), protons exhibit a
pronounced increase in net PKE, while neutrons show minimal or even attractive
contributions, particularly in the $^{16}$O+$^{208}$Pb interaction. Dynamically, the
primary distinction between symmetric and asymmetric encounters is the
activation of nucleon transfer channels in the latter, spurred by $N/Z$
equilibration across nuclei. This quick process, unfolding over approximately 1
zs, initiates upon nuclear contact \cite{simenel2020}. Expected are proton transfers from lighter
($N = Z$) to heavier ($N > Z$) nuclei, with neutrons moving oppositely, aligning
with experimental observations. The Pauli exclusion principle hinders protons
from occupying already filled states, potentially explaining the amplified net
PKE due to proton dynamics. Conversely, neutrons may transfer to vacant states
without contravening the Pauli principle, suggesting no significant PKE
increase. This hypothesis could be further explored using TDHF to
microscopically analyze the transfer's initial and final states, setting the
stage for future investigations.

\section{Summary}
The significance of the Pauli exclusion principle in formulating and refining
models to compute the interaction between two nuclei has been a focal point of
research for a long time. This interest stems from the observation that
the effects of antisymmetrization were often considered secondary in many
computational strategies. Moreover, there has been debate over the minimal
impact of such effects at the peak of the interaction barrier. Contrary to
initial beliefs, deviations from this assumption were first recognized in
alpha-nucleus potentials and later in scenarios involving higher bombardment
energies or energies significantly below the barrier. Attempts to address these
discrepancies have ranged from straightforward antisymmetrization of nuclear
states—requiring normalization due to the non-representation of lowest energy
configurations—to adopting empirically modified shallow potential models that
incorporate tailored potential wells.

Here, the novel DCFHF methodology has been applied to derive pure nucleus-nucleus
potentials in reactions involving $^{40,48}$Ca+$^{40,48}$Ca and
$^{16}$O+$^{208}$Pb. This technique comprehensively incorporates
antisymmetrization while optimizing the system's energy by adjusting orbitals
against a static density framework. A key feature of the DCFHF approach is its
precise adherence to the Pauli exclusion principle (at a mean-field
approximation level) without necessitating additional parameters beyond those
used in constructing the Skyrme EDF for the nuclear mean-field. Compared to
traditional FHF methods, which overlook the Pauli exclusion among nucleons from
different nuclei, DCFHF allows for the quantification of Pauli-induced nuclear
repulsion, effectively broadening and elevating the interaction barrier to
impede fusion below the barrier.

In our research, we've utilized the expression for Pauli kinetic energy derived
from nuclear localization function studies within density functional theory.
This reveals that repulsion predominantly arises in the fragment's connecting
region at distances matching the barrier's radius. Analyzing the Pauli kinetic
energy distribution further allows for the dissection of its proton and neutron
components, highlighting a dominant neutron effect within neutron-rich systems,
thereby proposing additional resistance to sub-barrier fusion in such contexts.

Explorations into the dynamic behaviors of Pauli kinetic energy using the
DC-TDHF method have unveiled that the system tends to navigate towards
minimizing Pauli repulsion near the barrier. Inside the barrier, though, dynamic
processes generally elevate the Pauli kinetic energy as nuclei combine and
nucleons undergo collectivization. Notably, the dynamical contributions of
protons and neutrons to the Pauli kinetic energy significantly diverge in the
examined asymmetric systems, interpreted as a result of multinucleon transfer
propelled by swift $N/Z$ balancing. The potential for states transfer with
favorable $Q$-values to mitigate Pauli repulsion highlights a complex interplay
that could enhance tunneling probabilities within the Coulomb barrier.

Looking ahead, investigations will focus on the role of Pauli repulsion in
mid-shell nuclei to understand better the influences of nuclear pairing and
deformation. Additionally, delving into the dynamics of single-particle states
involved in transfer mechanisms through microscopic analyses could offer
invaluable insights into the nuanced relationship between transfer processes and
Pauli kinetic energy.
A limitation of the current DC-TDHF approach is that it relies on TDHF trajectories above the Coulomb barrier. To investigate the role of dynamics on Pauli kinetic energy at sub-barrier energies, one would need to extend TDHF to account for many-body tunnelling, e.g., with imaginary time-dependent mean-field methods \cite{levit1980c,mcglynn2020}. \\

\begin{acknowledgement}
This work has been supported by the U.S. Department of Energy under award numbers DE-SC0013847
(Vanderbilt University) and DE-NA0004074 (NNSA, the Stewardship Science Academic Alliances program)
and by the Australian Research Council Discovery Project (project number DP190100256) funding schemes.
\end{acknowledgement}

%
\bibliography{VU_bibtex_master}

\end{document}